\documentclass[a4paper]{article}

\usepackage{INTERSPEECH2020,subfig}

\title{Length- and Noise-aware Training Techniques\\ for Short-utterance Speaker Recognition}
\name{Wenda Chen$^{1,}$\sthanks{*Equal Contribution}, Jonathan Huang$^{2,*}$\sthanks{\hspace{0.1cm} Work done while at Intel Labs}, Tobias Bocklet$^{1,3}$}
\address{
  $^1$Intel Labs\\
  $^2$Apple Inc.\\
  $^3$Technischen Hochschule Nürnberg}

\email{wenda.chen@intel.com, jjhuang@apple.com, tobias.bocklet@intel.com}

\begin{document}

\maketitle
%
\begin{abstract}
Speaker recognition performance has been greatly improved with the emergence of deep learning. Deep neural networks show the capacity to effectively deal with impacts of noise and reverberation, making them attractive to far-field speaker recognition systems. The x-vector framework is a popular choice for generating speaker embeddings in recent literature due to its robust training mechanism and excellent performance in various test sets. In this paper, we start with early work on including invariant representation learning (IRL) to the loss function and modify the approach with centroid alignment (CA) and length variability cost (LVC) techniques to further improve robustness in noisy, far-field applications. This work mainly focuses on improvements for short-duration test utterances (1-8s). We also present improved results on long-duration tasks. In addition, this work discusses a novel self-attention mechanism. On the VOiCES far-field corpus, the combination of the proposed techniques achieves relative improvements of $7.0\%$ for extremely short and $8.2\%$ for full-duration test utterances on equal error rate (EER) over our baseline system.

\end{abstract}
\noindent\textbf{Index Terms}: speaker recognition, invariant representation learning, centroid alignment, x-vector, far-field

\section{Introduction}

Speaker recognition system have been popularized in consumer devices such as smart speakers and smartphones.  In these use cases, the speech utterance of the user during an interaction with the device can be used to perform voice matching.  These use cases are particularly challenging with respect to channel degradation. Furthermore, the interactions tend to be short, making recognition even harder. In this paper, we propose techniques to address these issues. 
Recently, a successful application of deep neural networks to the domain of speaker recognition helped to improve the accuracy significantly \cite{Zhang2017-ETS}. Here, the network is trained end-to-end via triplet loss.
Another state of the art system is x-vector extraction \cite{Snyder18-XRD}. A frame-wise neural network is used as feature generator followed by a segment-based network using pooled frame-wise data.
Training is performed via cross-entropy loss on multiple thousand speakers. For recognition, speaker embeddings are extracted and compared with a similarity measure, e.g., cosine distance. 
Noise robustness is achieved by PLDA \cite{Ioffe06-PLDA} and simulation of noisy training data \cite{Meng14-NTF}. 

Recent research evaluated various extensions of the x-vector approach focusing also on higher noise robustness. The statistical pooling was substituted by attention mechanism \cite{wang2018attention,zhu2018self2,Han2019,an19deep} in order to learn the weighted pooling in a data-driven manner. 
Angular softmax \cite{huang2018angular} and large margin losses \cite{wang2018additive} showed in combination with noise-augmented training data very good results. 
\cite{ferrer20discriminative} presented a scoring approach mimicing the PLDA-scoring by discriminative training. 
\cite{Novoselov2019,gusev2020deep} added residual blocks in order to allow deeper networks. 
\cite{Huang2019,cai2020withinsample,liang2018learning} motivated loss functions in order to teach the network to learn an implicit cleaning of encodings.

Besides noise, the duration of utterances is a factor of high importance for the accuracy. 
\cite{Chakroun20-RFF} described the use of gammatone features in combination with i-vector on short utterances, \cite{Bhat17-DSE} trained deep convolutional networks specifically on short utterances.
\cite{Zhang2017-ETS} described the use of inception networks by transforming utterances to fixed length spectrograms and training via triplet loss.
\cite{Ji18-AET} used stacked gated recurrent units (GRU) to extract utterance-level features followed by residual convolution neural networks (ResCNNs) trained with speaker identity subspace loss that focuses on transforming utterances from the same speaker to the same subspace.
\cite{liu20text} used adversarial learning techniques to learn enhanced embeddings from short and long embedding pairs from the same speaker.
\cite{tawara20-frame} argued that pooling and phoneme-aware training is harmful, especially for short-utterance SID and used adversarial training to remove phoneme-variability from the speaker embeddings.

This work focuses on enhanced training techniques for x-vector embeddings with a strong focus on short-duration speech segments in heavy far-field and noisy conditions. 
We start with our previous work \cite{Huang2019} on adding additive margin softmax (AM-softmax) and Invariant Representation Learning (IRL) to an x-vector-based SID system. Due to recent improvements, we substitute the statistical pooling by an attention mechanism. We then modify the idea of IRL and introduce improved training techniques: Centroid Alignment (CA) and Length Variability Cost (LVC). These techniques address the variability in utterances due to channel noises and utterance duration, by making the embedding space less sensitive to such perturbations. CA modifies IRL by enforcing small distances of the training utterance to the average embedding (centroid). LVC tries to keep the distance between a speaker's two randomly-selected utterances with different lengths small. The proposed techniques show improvements on both short and full-length VOiCES test utterances which were collected in noisy and reverberant environments \cite{nandwana2019voices}. 
The rest of the paper is organized as follows:  Section~\ref{sec:motivation} provides the intuition behind our training techniques.  Section~\ref{sec:architectures} details the modifications we made to the x-vector architecture as well as our attention mechanism.  Section~\ref{sec:learning} describes the model training techniques.  Section~\ref{sec:results} shows the results and discussions of our experiments. We finish the paper with a short conclusion in Section~\ref{sec:conclusion}.

\section{Motivation}    
\label{sec:motivation}
\begin{figure}[!htbp]
\centering
\begin{tabular}{|c|c|}
\hline
[a]{\includegraphics[width=0.35\columnwidth,height=0.3\columnwidth]{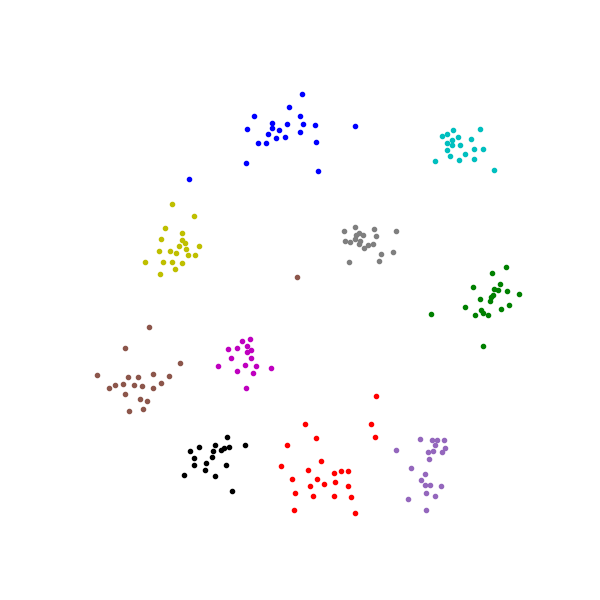}}
   &
[b]{\includegraphics[width=0.35\columnwidth,height=0.3\columnwidth]{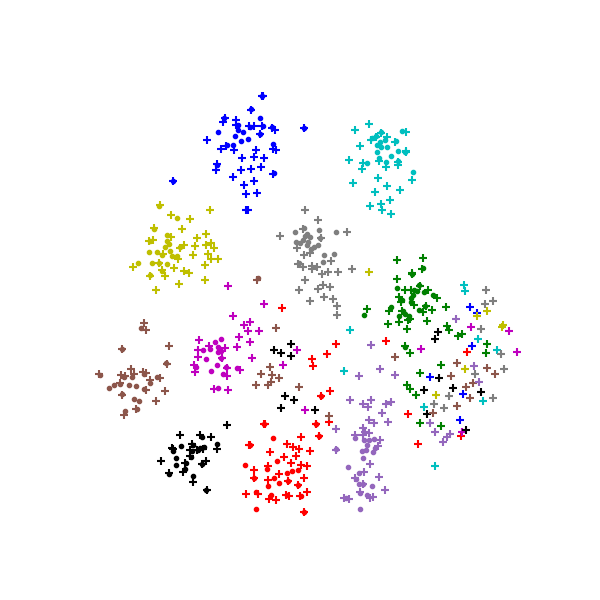}}
\\
\hline
[c]{\includegraphics[width=0.35\columnwidth,height=0.3\columnwidth]{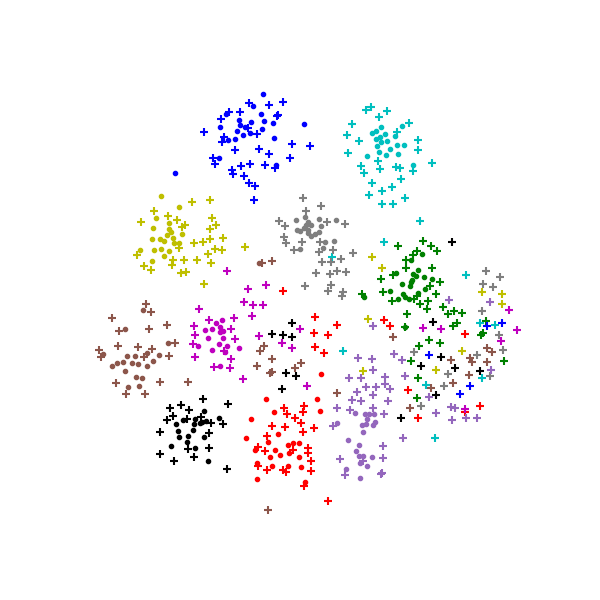}}
     &
[d]{\includegraphics[width=0.35\columnwidth,height=0.3\columnwidth]{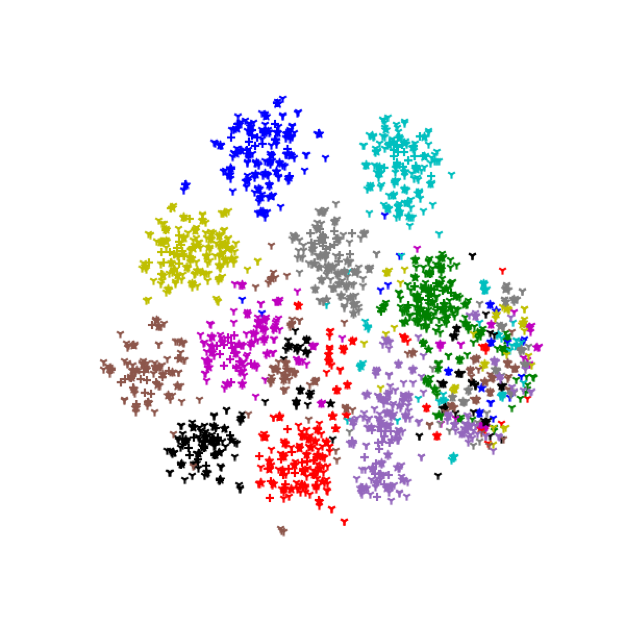}}
\\
\hline
\end{tabular}
 \caption{t-SNE visualization of embeddings for 10 speakers represented by distinct colors. [a]=full-length clean speech; [b]=full-length clean + noisy speech; [c]=full-length clean + short; [d]=all of the above.}
   \label{fig:centroid}
\end{figure}

The intuition behind our model training techniques can be best illustrated by visualizing the 2-D t-SNE plots of a well-trained speaker embedding (Fig.~\ref{fig:centroid}). The embedding was been generated from the baseline system described in Section~\ref{ssec:xvec}. The plots show embeddings extracted from random utterances of 10 speakers in LibriSpeech \cite{Pana15-LAA}. The full-length clean speech utterances shows distinct, well-separated clusters; no confusion with other speakers is visible. When noise is added to the full-length utterances [b], the speaker clusters are more blurred. Adding clean speech segments truncated to 2s [c] shows higher confusion than [b]. Combining noise and short duration amplifies the effect [d].  To also quantify the growing deviation, we calculated the average standard deviation of the x-vector embeddings (see line \emph{baseline} in \tablename~\ref{tab:example}): noisy embeddings show a higher deviation compared to clean, embeddings generated on short utterances have a higher deviation than noisy ones and noisy short further increases the standard deviation. Based on the simple observation that clean speech is `better' than noisy, and full-length is `better' than truncated, we hypothesize that we can use the `better' samples as an additional training target. This idea is the basis for our auxiliary training objectives:  for IRL, we align the noisy speech with its corresponding clean version; for LVC, we align a short duration speech with its full-length version; and for CA, we align the noisy or short-duration speech with the cluster centroid of clean full-length speech for each speaker.

\section{System Architectures}\label{sec:architectures}

\subsection{Baseline X-vector Architecture}
\label{ssec:xvec}
Although our contributions in attention and model training work for other embedding architectures, we show our findings using the popular x-vector architecture \cite{Snyder18-XRD}. 
The topology we used is a slight modification of the original x-vector description.  We chose 40-dimension log mel-filterbank as features. The features are mean-normalized on a 3-second sliding window. In \tablename~\ref{table:xvec_topo}, layers 1-6 are unmodified from the original paper.  We removed one of the dense layers before the output, and extract the embedding at layer 7 with 256 dimensions. Our experiments have shown consistently better results with these modifications.

\begin{table}[!htbp]
\caption{Baseline x-vector system configuration, with $N$ speakers in model training. }
\label{table:xvec_topo}
\small
\centering
\setlength{\tabcolsep}{4.5pt}
\vspace{3pt}
\begin{tabular}{|c|c|c|c|}
    \hline
    Layer & layer type & layer context & output dimension  \\
    \hline
        1  & TDNN-ReLU & [t-2,t+2]  & $512$  \\
        2  & TDNN-ReLU & \{t-2,t,t+2\} & $512$  \\
        3  & TDNN-ReLU & \{t-2,t,t+2\} & $512$  \\
        4  & Dense-ReLU & \{t\}         & $512$  \\
        5  & Dense-ReLU & \{t\}         & $1500$  \\
        6  & Stats pool & [0,T] & $3000$  \\
        7 & Dense-ReLU & [0,T] & $256$  \\
        8 & Dense & [0,T] & N \\
    \hline
    \end{tabular}

\end{table}

\subsection{Multi-headed Self-attention for X-vector}
\label{ssec:attention}
Attention mechanisms have been proposed in several previous studies \cite{wang2018attention,zhu2018self2,rahman2018attention} to give different weightings to the frames within an speech utterance.  In the context of the x-vector embedding, the stats pooling layer is replaced with an attentive pooling layer.  The input to the attentive pooling is the hidden representation of layer 5, which we define here as $\mathbf{h}_t$ at time step $t$.  In a departure from previous literature, we compute different attention heads for parts of $\mathbf{h}_t$.  Our hypothesis is that the dimensions in this hidden representation correspond to different aspects of the signal which should not be treated with the same weighting.  Concretely, we break the vector $\mathbf{h}_t$ into $K$ contiguous, equal-length, smaller sub-vectors $\mathbf{h}_{t,k}$, where $k=1,...,K$ is the sub-vector index.  The attention weight corresponding to these sub-vectors can be computed by
\begin{equation} \label{eq:att_weight}
	e_{t,k} = f(\mathbf{w}^{T}_{k}\mathbf{h}_t+b_k),
\end{equation}
where $\mathbf{W}_k \in \mathbb{R}^{1500 \times 1500/K}$ and  $b_k \in \mathbb{R}^{1500/K \times 1}$ are the trainable weight matrix and bias, respectively, and $f$ is the non-linear activation.  We found that the Sigmoid activation here gives the best performance.  The frames of the attention weights are normalized by a Softmax 
\begin{equation} \label{eq:att_softmax}
	\alpha_{t,k} = \frac{exp(e_{t,k})}{\sum_{t=1}^{T}exp(e_{t,k})}
\end{equation}
The attentive mean and standard deviation pooling for each sub-vector across the entire utterances for $t=1,...,T$ time steps is
\begin{equation} \label{eq:mean_pool}
	\boldsymbol\mu_k = \sum_{t=1}^{K} \alpha_{t,k} \mathbf{h}_{t,k},
\end{equation}
and 
\begin{equation} \label{eq:std_pool}
	\boldsymbol\sigma_k = \sqrt{\sum_{t=1}^{K} \alpha_{t,k} 
	\mathbf{h}_{t,k} \otimes \mathbf{h}_{t,k}
	- \boldsymbol\mu_k \otimes \boldsymbol\mu_k
	},
\end{equation}
respectively. $\otimes$ denotes the element-wise multiply operation.  Finally, the output of the attentive pooling is formed by a concatenation of the results from \ref{eq:mean_pool} and \ref{eq:std_pool}, for all $k$,
\begin{equation} \label{eq:concat}
	\mathbf{z} = [\boldsymbol\mu_1^T,...,\boldsymbol\mu_K^T,
	\boldsymbol\sigma_1^T,...,\boldsymbol\sigma_K^T]^T
\end{equation}
as a 3000-dimension vector for the utterance.

\section{Training Techniques}
\label{sec:learning}

\subsection{Baseline Training}
\label{ssec:amsoftmax}
In the original x-vector paper, the output of the last layer with N speaker labels is fed into a Softmax loss.  Because Softmax is specifically designed for classification, others focused on improving the speaker recognition objective with other loss functions \cite{Zhang2017-ETS, li2017deep} or even completely different training infrastructure \cite{heigold2016end}. In our experiments, the use of Additive Margin Softmax (AM-softmax) loss \cite{wang2018additive} was consistently superior to basic Softmax training in the far-field test set. The systems trained with AM-softmax did not benefit from having a PLDA backend \cite{Ioffe06-PLDA}. We use cosine similarity as the scoring mechanism. This then reflects our baseline system.

For the proposed learning techniques which are describing next, the weights are initialized with the weights of a baseline systems. We used two different methodologies to train our baseline systems: 1. with fixed length 8s long utterances, 2. with utterances of variable length between 0.5 and 8.5s. In the result section, these variants are differentiated by the suffix \emph{-long} for the former and \emph{-varied} for the later variant.

\subsection{Invariant Representation Learning}

\begin{figure}[!htbp]
	\includegraphics[width=0.95\linewidth]{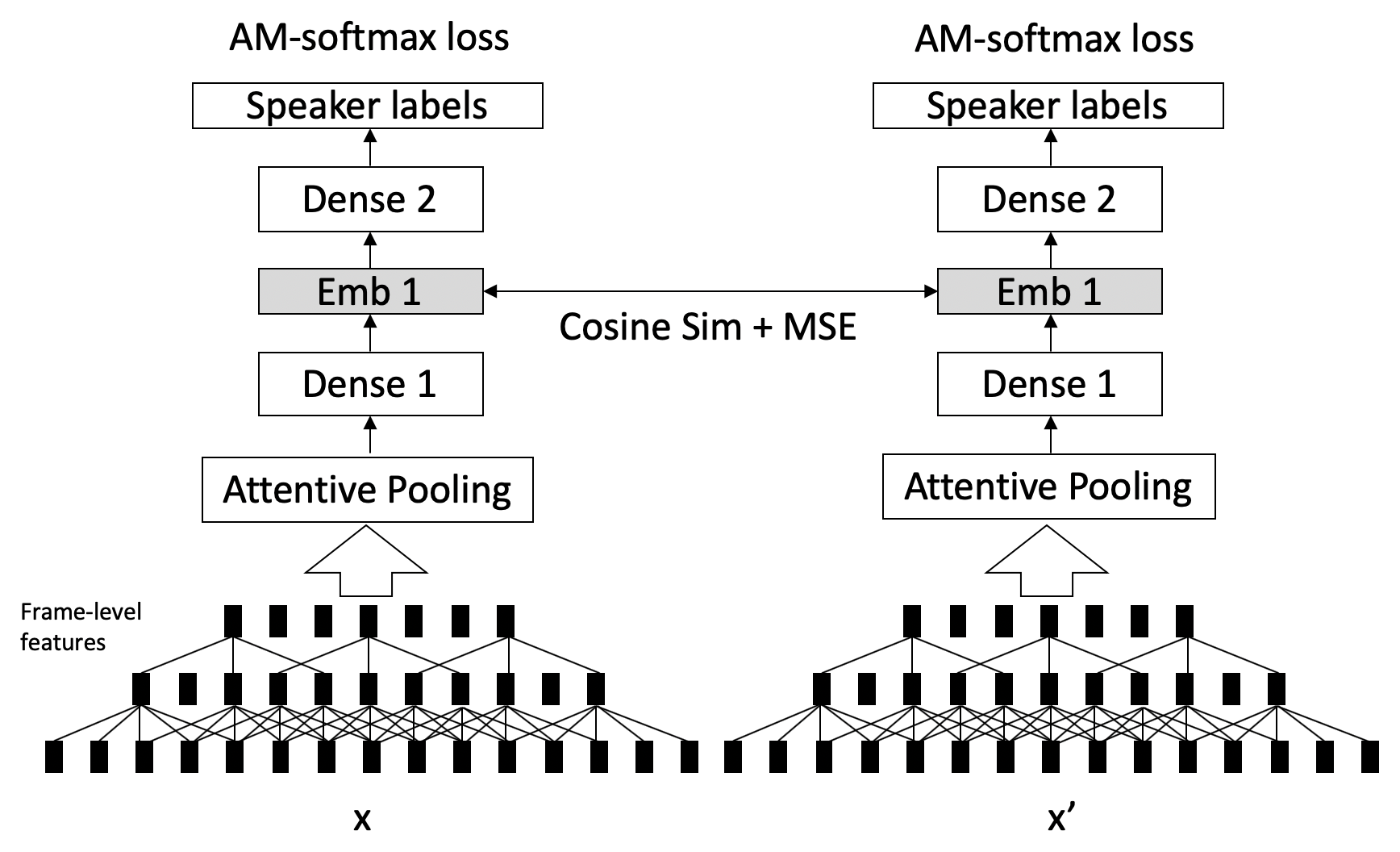}
	\caption{IRL utilizes parallel clean and noisy speech in each training iteration to train the weights of a single network.  The main loss is the AM-softmax, but auxiliary loss functions are introduced to explicitly align the embeddings.}
	\label{fig:irl}
\end{figure}

Fig.~\ref{fig:irl} shows the IRL training procedure.  At each training iteration, clean ($x$) and noisy ($x'$) features of the same utterance are fed into the network one after another, resulting the computation of two AM-softmax loss values.  At the layer where the speaker embedding is extracted, we align the two sides by imposing cosine similarity and MSE loss.  Mathematically, the combined loss is sumarized as 
\begin{equation} \label{eq:irl}
    L_{IRL}(x,x')=L_{AM}(x)+\alpha L_{AM}(x')-\gamma L_{cos}(x,x')+\lambda L_{2}(x,x')
\end{equation}
where $L_{AM}$ is the categorical loss, $L_{cos}$ is the cosine similarity (\ref{eq:Lcos}), and $L_2$ is the means square loss (\ref{eq:L2})
\begin{equation} \label{eq:Lcos}
	L_{cos}(x, x') =   
	\frac{\phi(x) \cdot \phi(x')}{\|\phi(x)\| \|\phi(x')\|}
\end{equation}
\begin{equation} \label{eq:L2}
	L_{2}(x, x') =  (\phi(x)-\phi(x'))^2
\end{equation}

The embedding layer representation is denoted by $\phi_l$.  The parameters $\alpha$, $\gamma$, and $\lambda$ control the contributions from the losses.  This training tries to learn a representation invariant to noise. For IRL we achieved best results when initializing with a baseline trained on fixed 8s utterances.

\subsection{Length Variability Cost}
LVC is in fact a special case of IRL.  Here, $x$ is a long-duration utterance (i.e. 8s), and $x'$ is a truncated version of the same utterance (i.e. randomly chosen length between 0.5-8.5s). This training tries to learn a representation invariant to utterance duration. $\alpha =1$ for LVC and IRL in Eq.~\ref{eq:irl}. For LVC we achieved best results when initializing with a baseline trained by the fixed length methodology.

\subsection{Centroid Alignment}
The intuition behind CA is that the clean full-length utterances, and especially their centroid of each speaker, is a desirable training target in the embedding space. After each epoch of training, we compute the embeddings for all clean full-length segments, and average the length-normalized embeddings for each speaker label to form the centroids.  For the subsequent epoch, the speaker centroid is the target to do alignment against each training sample.  Concretely, for Eq.~\ref{eq:irl}, we compute the AM-softmax for $x$ only, and set $\alpha$ to zero because we do not use training pairs.  Instead of doing alignment with another training sample, we use the centroid of speaker for $x$ as the alignment target $x'$. 

\section{Experimental Results}\label{sec:results}

\begin{figure}[tp]
  \centering
  \includegraphics[width=0.95\linewidth]{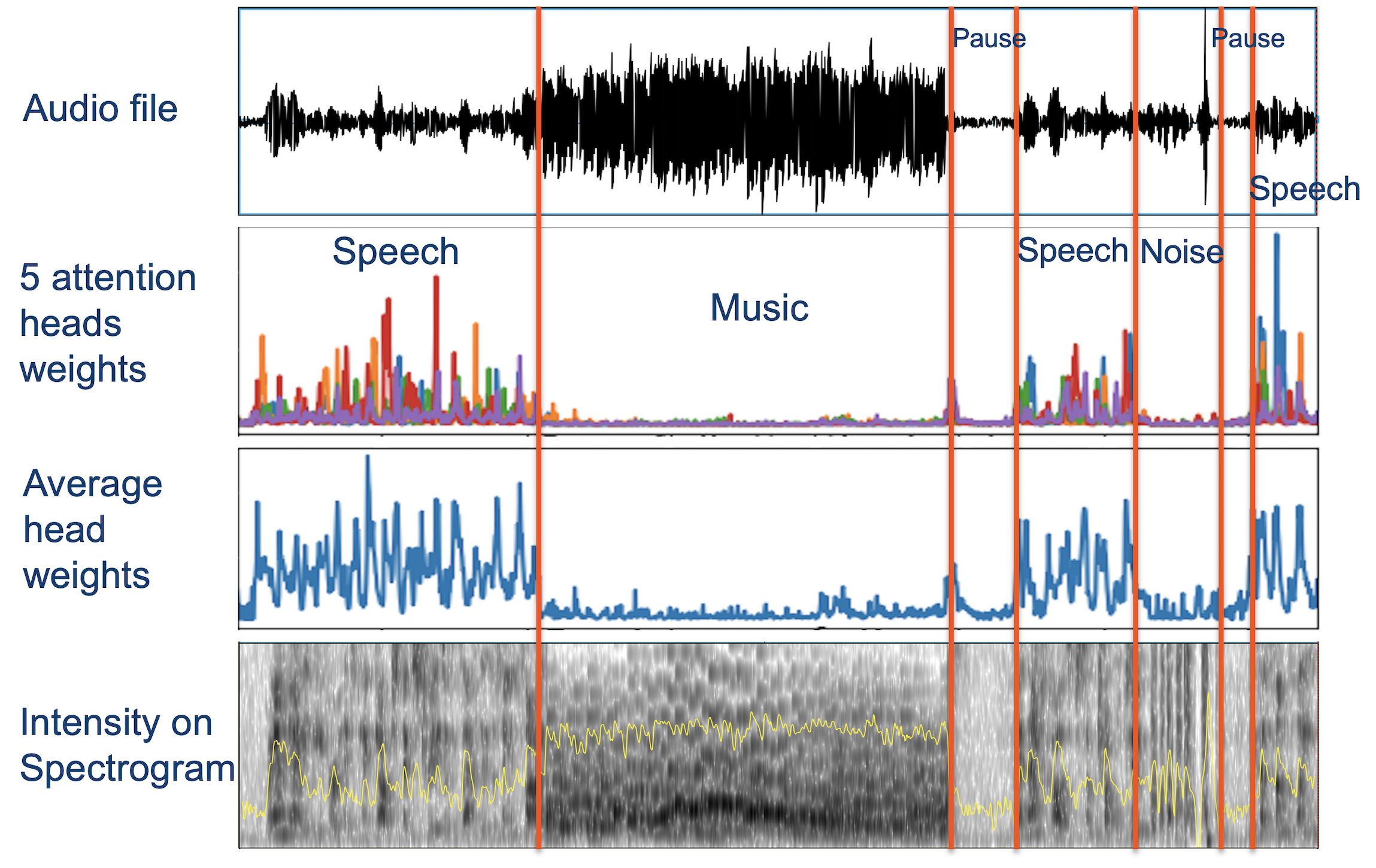}
  \caption{Attention weights are used as the speech activity detectors.}
  \label{fig:attention}
\end{figure}

    \begin{table}[bp]
  \caption{EER (in \%) and minDCF results on VOiCEs Dev and Eval achieved with AM-softmax training on 8s speech segments with and without attention using different x-vector configurations.}
  \label{tab:attResult}
  \centering
  \begin{tabular}{r r r r r}
    \toprule
    &\multicolumn{2}{c}{Dev}& \multicolumn{2}{c}{Eval}\\
    System  & EER &minDCF& EER &minDCF \\
    \midrule
    x-vector & 1.78&0.184&5.69&0.374\\
    x-vector-att & 1.59&0.190&5.61&0.363\\
    ext-x-vector & 1.32&0.149&5.02&0.314\\
    ext-x-vector-att& 1.31&0.128&4.99&0.308\\
    \bottomrule
  \end{tabular}
\end{table}

\begin{table*}[!tp]
  \caption{EER (in \%) and minDCF results on VOiCES Dev for the x-vector-att topology trained with various loss functions(AMSM: AM-softmax, LVC, CA, IRL). \emph{-long} refers to fix 8s segment training; \emph{-varied} refers to training on variable length speech segments (0.5-8.5s). Best result per eval condition are marked in bold.}
  \label{tab:resultBig}
  \centering
  \begin{tabular}{r r r r r r r r r r r}
    \toprule
    Test Duration & \multicolumn{2}{c}{AMSM-long} & \multicolumn{2}{c}{AMSM-varied} & 
    \multicolumn{2}{c}{LVC}  & \multicolumn{2}{c}{CA}  & \multicolumn{2}{c}{IRL}
 \\
    \midrule
  1s dev   &	15.03 & 0.881  &	12.71 &0.842  &	13.68&0.857  &	\textbf{11.82}&\textbf{0.828}  &14.73&0.874  \\
  2s dev   &	7.77 & 0.658  &	6.72&0.638  &	7.13&0.625  &	\textbf{6.59} & \textbf{0.618}  &7.52&0.650  \\
      4s dev   &	3.81&0.431  &	4.02&0.457  &	\textbf{3.68}&\textbf{0.417}  &	3.70&0.437  &3.83&0.426    \\
          8s dev   &	2.38&0.286  &	2.50&0.322  &	\textbf{2.09}&\textbf{0.270}  &2.46&0.308    &2.21&0.263.    \\
       Full dev   &	1.59&0.190  &	1.89&0.233  &	1.50&0.179  &	1.78&0.228  &\textbf{1.46}&\textbf{0.171} \\
    \bottomrule
  \end{tabular}
\end{table*}

\subsection{Experimental Settings}
Our systems are trained on the Voxceleb 1 and 2 corpora \cite{Nagrani20-VLS}. The training material is prepared by applying 10x data augmentation. For each augmented speech file we convolve a randomly chosen room impulse response (RIR) from 100 artificially generated RIRs by Pyroomacoustics \cite{scheibler2018pyroomacoustics} and 100 randomly selected real RIRs from the Aachen Impulse Response Database \cite{jeub2009binaural}. Afterwards, the data is mixed in with randomly chosen clips from Google's Audioset under Creative Commons \cite{gemmeke2017audio}. The SNR for mixing was uniformly distributed between 0 and 18\,dB.  
We evaluate the proposed loss functions on the VOiCES Dev and Eval data \cite{nandwana2019voices} and compare against a baseline system trained with AM-softmax loss. 
We found optimal weighting factors $\gamma$ and $\lambda$ for CA  with $\gamma=0.5$ and $\lambda=0.01$, meaning the cosine loss is weighted significantly higher. For both LVC and IRL ideal values have been found with $\gamma=\lambda=0.5$. We see very consistent results between Dev and Eval. For the sake of readability we present numbers on VOiCES dev EER and  minDCF ($P_{tar}=0.01$) metrics.

     \begin{table}[bp]
  \caption{Average standard deviation of embeddings for the 10 speakers and conditions described in Fig.~\ref{fig:centroid} before and after CA and LVC techniques.}
  \label{tab:example}
  \centering
  \begin{tabular}{r r r r r}
    \toprule
     & Clean & Noisy & Short & Noisy \& short 
 \\
    \midrule
      Baseline   &	0.0202  &	0.0284  &	0.0367 & 0.0394      \\
      After CA   &	0.0187  &	0.0282  &	0.0358 & 0.0386      \\
      After LVC  &	0.0189  &	0.0283  &	0.0349 & 0.0376      \\
    \bottomrule
  \end{tabular}
\end{table}

\subsection{Self-attention Model}
Fig.~\ref{fig:attention} visualizes the behaviour of the attention mechanism on an audio segment containing speech, music, and silence.  If we look at 5 randomly chosen attention heads, their weights are different in the speech segments, indicating they are learning different aspects of the signal.  The average of the attention weights across all heads shows that the silence and noisy segments of the audio are suppressed compared to the speech.  Empirically we saw improvements in increasing the number of attention heads to 100, but not beyond.  We use 100 heads for the results here. In \tablename~\ref{tab:attResult} the positive influence of attention is visible in the recognition rates. \emph{x-vector} denotes our baseline system with the topology shown in \tablename~\ref{table:xvec_topo}. For the sake of completeness, we also show results of \emph{ext-x-vector} with and without attention. \emph{ext-x-vector} is an extended x-vector system as defined in \cite{Snyder2019}. The temporal context is slightly
wider for the frame-level layers and dense layers are added in between the T-DNN layers. The x-vector variants with attention are consistently better than the non-attention variants. The extended variant  significantly outperforms the baseline. For the sake of less training time and limited model size and power, we decided to use the smaller topology \emph{x-vector-att} for the evaluation of the proposed loss functions.

\subsection{LVC and CA loss}
The models are evaluated on various speech durations: 1s, 2s, 4s, 8s, and full (as in VOiCES). Results are summarized in \tablename~\ref{tab:resultBig}. For AM-softmax we report results on the training methodologies \emph{-long} (8s training samples) and \emph{-varied} (0.5-8.5s training samples). For AM-softmax, the results on short utterances (1-4s) are superior when trained with variable length methodology. On 8s and full duration, training via long methodology achieved the better results. This confirms the intuition: training a system on long utterances gives better results on longer test utterances. 
CA shows an advantage in extremely short test utterances (1s and 2s). CA tries to bring short utterances closer to the speaker centroid which is estimated on 8s long utterances and thus helps to stabilize results on short utterances. On 1s test utterances the EER is reduced by 7.0~\% compared to AMSM-varied.  
LVC seems especially helpful for 4s and 8s durations.  LVC tries to diminish the distance between long, i.e., 8s and randomly a shorter version (randomly chosen 0.5-8.5s) of the same utterances. 
Thus, LVC helps to stabilize the training process by removing general speaker variability, which is also visible in \tablename~\ref{tab:example}. On 8s segments, LVC improves the EER result by 12.2\,\% compared to AMSM-long. IRL works best in the full length condition (8s pairs of utterances from the same speaker are aligned) and is close to LVC in the 8s test condition. Trying to tie a clean and noisy utterances of the same size close together, is especially helpful for longer duration utterances: IRL achieves an EER reduction of 8.2\,\% on the full duration utterances compared to AMSM-long.

\tablename~\ref{tab:example} shows the average standard deviations of embedding for the 10 speakers  plotted in \figurename~\ref{fig:centroid} in clean, noisy and short scenarios. Compared to the AM-softmax baseline, CA and LVC help to reduce the variability especially for short and combined (noisy and short) scenarios. For the baseline, the average standard deviations increases for noisy, short and the combination. Applying CA and LVC, brings noisy and short utterances closer to the centroids and removes variation within utterances of one and the same speaker. Hence, making the speaker separation more accurate. While the clean cases also get closer to their cluster centroids, the average relative improvement rate is especially high for the short and noisy cases at 4.6\%. 


\section{Conclusion} \label{sec:conclusion}
In this paper, we proposed to extend the x-vector based SID with self-attention pooling and apply new loss functions based on IRL, CA and LVC. These techniques are targeted for improving speaker recognition on short utterances in reverberant and noisy conditions. Both the proposed attention mechanism and the improved loss functions show a reduction in error rate. Our experiments and analysis proved that the centroids and embedding length variations are good regularization references for the AM-softmax losses. Combining the approaches leads to an EER reduction of 7.0\,\% on 1s and 8.2\,\% on full duration utterances.
\bibliographystyle{IEEEtran}

\bibliography{mybib}

\end{document}